\documentclass[aps,prb,superscriptaddress,twocolumn,showpacs,floatfix]{revtex4-1}

\usepackage{graphicx}
\usepackage{dcolumn}
\usepackage{bm}
\usepackage{color}
\usepackage[colorlinks=true,urlcolor=blue,citecolor=blue, linkcolor = blue]{hyperref}
\usepackage{amssymb,ulem,amsmath}
\def\be{\begin{equation}}
\def\ee{\end{equation}}
\def\ba{\begin{array}{lll}}
\def\ea{\end{array}}
\def\ber{\begin{eqnarray}}
\def\eer{\end{eqnarray}}
\begin{document}
\title{Quantum Breathing of an Impurity in a One-dimensional Bath of Interacting Bosons}
\author{Sebastiano Peotta} 
\email{s.peotta@sns.it}
\affiliation{NEST, Scuola Normale Superiore
  and Istituto Nanoscienze-CNR, I-56126 Pisa, Italy}
\author{Davide Rossini} \affiliation{NEST, Scuola Normale Superiore
  and Istituto Nanoscienze-CNR, I-56126 Pisa, Italy}
\author{Marco Polini} \affiliation{NEST, Istituto Nanoscienze-CNR and
  Scuola Normale Superiore, I-56126 Pisa, Italy}
\author{Francesco Minardi}
\affiliation{LENS-European Laboratory for Non-Linear 
Spectroscopy and Dipartimento di Fisica, Universit\`a di Firenze, 
via N. Carrara 1, IT-50019 Sesto Fiorentino-Firenze, Italy}
\affiliation{CNR-INO, via G. Sansone 1, IT-50019 Sesto
  Fiorentino-Firenze, Italy}
\author{Rosario Fazio} \affiliation{NEST, Scuola Normale Superiore and
  Istituto Nanoscienze-CNR, I-56126 Pisa, Italy}
\begin{abstract}
  By means of time-dependent density-matrix renormalization-group
  (TDMRG) we are able to follow the real-time dynamics of a single
  impurity embedded in a one-dimensional bath of interacting
  bosons. We focus on the impurity breathing mode, which is found to
  be well-described by a single oscillation frequency and a damping
  rate.  If the impurity is very weakly coupled to the bath, a
  Luttinger-liquid description is valid and the impurity suffers an
  Abraham-Lorentz radiation-reaction friction. For a large portion of
  the explored parameter space, the TDMRG results fall well beyond the
  Luttinger-liquid paradigm.
\end{abstract}
\maketitle

\section{Introduction} 
The dynamics of impurities jiggling in
classical and quantum liquids has tantalized many generations of
physicists since the early studies on Brownian motion.  In particular,
the interaction of a quantum system with an external environment
strongly affects its dynamics~\cite{weiss,gardiner,caldeiraleggett}.
Because of this coupling, the motion of a quantum particle is
characterized by a renormalized mass, decoherence, and damping.
Polarons~\cite{feynmann}, originally studied in the context of
slow-moving electrons in ionic crystals, and impurities in
$^3$He~\cite{ketterson} are two prototypical examples in which the
bath is bosonic and fermionic, respectively.  These problems
have been at the center of great interest for many decades in condensed
matter physics.

Recent advances in the field of cold atomic
gases~\cite{reviewscoldatoms} have made it possible to observe and
study these phenomena from a different perspective and hence to
disclose new aspects not addressed so far. It is indeed possible to
accurately tune the coupling between a quantum particle and the bath
and to modify the many-body nature of the bath itself. Furthermore,
the dynamics of the dressed particle can be studied in real time, thus
giving direct access to both mass renormalization and damping. This
problem becomes of particular relevance if the bath is 
one-dimensional (1D). In this case interactions strongly affect
the excitation spectrum of the bath~\cite{giamarchi} and therefore the
effective dynamics of the coupled system. 

The dynamics of impurities in cold atomic gases has attracted a great
deal of experimental~\cite{expimpurities} and
theoretical~\cite{thimpurities} attention in recent years. In
particular, Catani {\it et al}.~\cite{Minardi:2012} have recently
studied experimentally the dynamics of K atoms (the ``impurities")
coupled to a bath of Rb atoms (the ``environment") confined in 1D
``atomic wires". Motivated by Ref.~\onlinecite{Minardi:2012} we perform a
time-dependent density-matrix renormalization group (TDMRG)~\cite{tdmrg} 
study of the dynamics of the breathing mode in a 1D bath of interacting bosons.

At a first sight one might think that the problem under consideration
reduces to the study of a particle coupled to a Luttinger
liquid~\cite{LLbath}.  As we will discuss in the remainder of this
work, it turns out that the nonequilibrium impurity dynamics eludes
this type of description. This is the reason why we choose to tackle
the problem with an essentially exact numerical method. 
Several features of the experimental data in Ref.~\onlinecite{Minardi:2012}
are also seen in our simulations. As we will discuss in the
conclusions, however, a detailed quantitative account of the data in Ref.~\onlinecite{Minardi:2012} 
may require additional ingredients and is outside the scope of the present work.
Here, we highlight a number of distinct signatures of the impact of
interactions on the breathing motion of an impurity, which are
amenable to future experimental testing.


%
\begin{figure}[t]
\begin{center}
\includegraphics[width=1.0\linewidth]{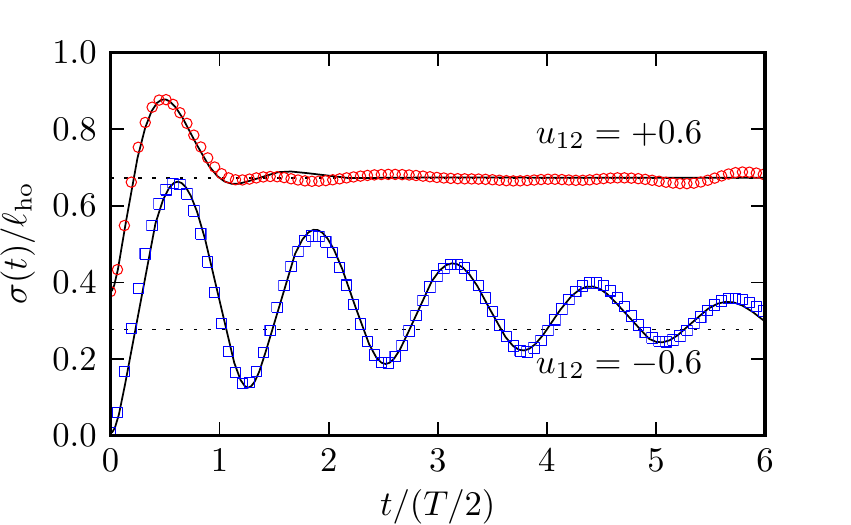} 
\caption{(Color online) The width of the impurity breathing mode
  $\sigma(t)$, in units of $\ell_{\rm ho} = (J_2/V_2)^{1/4}\delta$ ($\delta$ is 
  the lattice spacing), is plotted as a function of time $t$ (in units of
  $T/2 = \pi/\omega_2$). Data labeled by empty symbols represent TDMRG
  results corresponding to two opposite values of the impurity-bath
  coupling constant $u_{12}$. The strength of interactions in the bath
  has been fixed to $u_1=1$.  The data for $u_{12} = -0.6$ (squares)
  are shifted downward by $0.37~\ell_{\rm ho}$. The black solid lines
  are fits to the TDMRG data based on
  Eq.~(\ref{eq:motion_final}).\label{fig:one}}
\end{center}
\end{figure}

\section{Model Hamiltonian and impurity breathing mode}
\label{sec:model}
We consider a
1D bath of interacting bosons coupled to a single impurity confined in
a harmonic potential. The bath is modeled by a Bose-Hubbard
Hamiltonian with hopping $J_1$ and on-site repulsion $U_1>0$:
\be\label{eq:bath} {\hat {\cal H}}_{\rm B} =- J_1\sum_i ({\hat
  b}^\dagger_{i}{\hat b}_{i+1}+ {\rm H.c.}) + U_1\sum_i {\hat n}^2_{i}
+ \sum_iW_i{\hat n}_i~.  \ee
  Here ${\hat b}^\dagger_i$ (${\hat
  b}_i$) is a standard bosonic creation (annihilation) operator on the
$i$-th site. To avoid spurious effects due to quantum confinement along the 
1D system, we consider a nearly-homogeneous bath:
the external confining potential $W_i$ is zero in a
large region in the middle of the chain ($1 \leq i \leq L$) and raises
smoothly at the edges~(see Sec.~\ref{appendix:mapping}).  The local density
$\langle\hat n_i \rangle= \langle\hat b_i^\dagger \hat b_i\rangle$ is
thus essentially constant in a region of length $\sim 2L/3$.  In
(almost all) the results shown below we fix the number of particles in
the bath to $N_{\rm bath} = 22$ and distribute them over $L=250$
sites, thus keeping the average density to a small value, $\langle\hat
n_i \rangle\lesssim 0.1$.  For this choice of parameters the lattice
is irrelevant and the model (\ref{eq:bath}) is ideally suited to describe a
continuum.  Indeed, in the low-density limit, Eq.~(\ref{eq:bath}) reduces 
to the Lieb-Liniger model~\cite{cazalilla_rmp_2011} (a mapping between the 
coupling constants of the two models is summarized in Sec.~\ref{appendix:mapping}).
The Hamiltonian describing the impurity dynamics is
\begin{equation}\label{eq:impurity}
{\hat {\cal H}}_{\rm I}(t) = - J_2\sum_i\big({\hat a}^\dagger_{i}{\hat a}_{i+1}+ {\rm H.c.}\big) 
+V_2(t)\sum_{i}{\bar i}^2{\hat N}_i~,
\end{equation}
with ${\hat N}_i = {\hat a}^{\dagger}_{i}{\hat a}_i$ the impurity
density operator and ${\bar i} = i - i_0$. Eq.~(\ref{eq:impurity})
includes a kinetic term and a harmonic potential centered at $i_0 =
(L+1)/2$, whose strength $V_2(t)$ depends on time, mimicking the
quench performed in the experimental study of Catani {\it et
  al.}~\cite{Minardi:2012}. Because the on-site impurity density
$\langle \hat N_i \rangle \lesssim 0.15$ is low, also in this case the
model (\ref{eq:impurity}) well describes the corresponding continuum
Hamiltonian (Sec.~\ref{appendix:mapping}).  In this work we have fixed $J_2/J_1=2$ to
take into account the mass imbalance between the impurity and bath
atoms.  For future purposes we introduce $\omega(t) = 2\sqrt{V_2(t)
  J_2}/\hbar$, with $\omega_1 = \omega(t<0)$ and $\omega_2 =
\omega(t\geq 0)$, {\it i.e.} the harmonic-potential frequencies before
and after the quench, respectively.  The full time-dependent
Hamiltonian ${\hat {\cal H}}(t) = {\hat {\cal H}}_{\rm B} + {\hat
  {\cal H}}_{\rm I}(t) + {\hat {\cal H}}_{\rm coupl}$ contains a
further density-density coupling between bath and impurity ${\hat
  {\cal H}}_{\rm coupl} = U_{12}\sum_i {\hat n}_i{\hat N}_i$.

The quench in $\omega(t)$ excites the impurity {\it breathing} mode (BM),
{\it i.e.} a mode in which the width $\sigma(t)\equiv \big[\sum_i {\bar i}^2 \langle {\hat N}_i(t)\rangle
\big]^{1/2}$, associated with the
impurity density $\langle {\hat N}_i(t)\rangle$, oscillates in
time~\cite{footnotetime}.  This quantity is evaluated with the TDMRG.

\section{Numerical results} \label{sec:numerical} In Fig.~\ref{fig:one} we illustrate the
time evolution of the impurity width $\sigma(t)$ dictated by ${\hat
  {\cal H}}(t)$~\cite{initialcondition}. Different sets of data refer
to two values of the impurity-bath interaction $u_{12} =
U_{12}/J_2$. Time $t$ is measured in units of $T/2$, where
$T=2\pi/\omega_2$ is the period set by the harmonic-confinement
frequency $\omega(t)$ after the quench. The TDMRG results (empty
symbols) have been obtained by setting $u_1 = U_1/J_2 = 1$ and
$V_2(t)/J_2 = 10^{-3}$ for $t<0$ and $10^{-4}$ for $t\geq
0$~\cite{timescales}.

The black solid lines are fits to the TDMRG data based on the following
expression:
\be\label{eq:motion_final} \frac{\sigma^2(t)}{\sigma^2(+\infty)} = 1+
\frac{e^{-2\Gamma t}}{\cos^2(\phi)}\sum_{i=x,p}\Delta_{i}
\cos^2\left(t\sqrt{\Omega^2-\Gamma^2}-\theta_i\right) ~, \ee
where $\phi = \arccos(\sqrt{1-\Gamma^2/\Omega^2})$, $\theta_x=\phi$,
$\theta_p=\pi/2$, $\Delta_x = [\sigma(0)/\sigma(+\infty)]^2 -1$, and
$\Delta_p =
[\sigma(+\infty)/\sigma(0)]^2-1$. Eq.~(\ref{eq:motion_final}) is the
prediction for the BM width obtained by solving a quantum Langevin
equation for the impurity position operator ${\hat X}(t)$ in the
presence of Ohmic damping and a random Gaussian force with colored
spectrum (see Sec~\ref{appendix:quantum_breathing}):
\be 
\partial^2_t{\hat X}(t)+ 2 \Gamma \partial_t{\hat X}(t) +
\Omega^2{\hat X}(t) = {\hat \xi}(t)~.  
\ee
The three parameters $\sigma(+\infty)$, $\Omega$, and $\Gamma$ (respectively the asymptotic width at long times,
the frequency of the breathing oscillations, and the
friction coefficient) have been used to fit the data.  The
initial width $\sigma(0)$ is extracted from numerical data for the
ground-state width at $t< 0$. Note that, in the limit in which the
impurity-bath interaction is switched off ($u_{12} =0$), $\sigma(t)$
must oscillate at the frequency $2\omega_2$, since only even states
(under exchange $x \leftrightarrow -x$) of the harmonic-oscillator
potential are involved in the time evolution of a symmetric mode (like
the BM). This implies $\Omega \to \omega_2$ in the limit
$u_{12} \to 0$, where Eq.~(\ref{eq:motion_final}) reproduces the exact
non-interacting dynamics.

\begin{figure}[t]
\begin{center}
\includegraphics[width=1.0\linewidth]{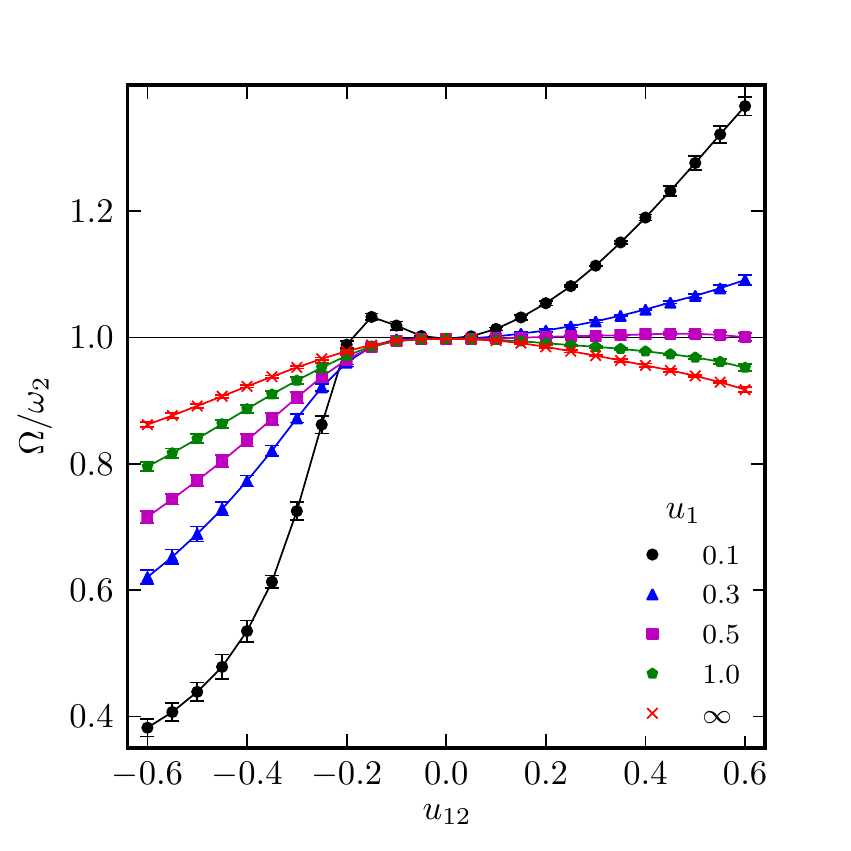} 
\caption{(Color online) The oscillation frequency of the breathing
  mode (in units of $\omega_2$) as a function of the impurity-bath
  Lieb-Liniger parameter $u_{12}$, for different values of the
  bath dimensionless coupling constant $u_1$. All the data in
  this figure have been obtained by setting $V_2(t)/J_2 = 10^{-3}$ for
  $t<0$ and $10^{-4}$ for $t\geq 0$. Error bars refer to the fitting procedure. 
  Solid lines are guides to the eye.\label{fig:two}}
\end{center}
\end{figure}

In Fig.~\ref{fig:two} we plot the values of the frequency $\Omega$ as
a function of $u_{12}$ and for different values of $u_1$. Several
features of the data in this figure are worth highlighting: i) the
behavior of $\Omega$ is dramatically different when the sign of
interactions is switched from attractive to repulsive, except at weak
coupling, for a tiny region of small $u_{12}$ values; ii) the behavior
becomes more symmetric with respect to the sign of $u_{12}$ as the
bath is driven deeper into the Tonks-Girardeau (TG) regime, {\it
  i.e.}, for $u_1 \to \infty$ (in passing, we notice that our results
in this limit are relevant in the context of the so-called ``Fermi
polaron" problem~\cite{fermipolaron}); iii) the renormalization of  the frequency 
$\Omega$ is reduced on increasing  the strength of repulsive interactions in the
bath $u_1$. It becomes almost independent of $u_{12}$ for  $u_1 \gtrsim 0.5$.

\begin{figure}[t]
\begin{center}
\includegraphics[width=1.0\linewidth]{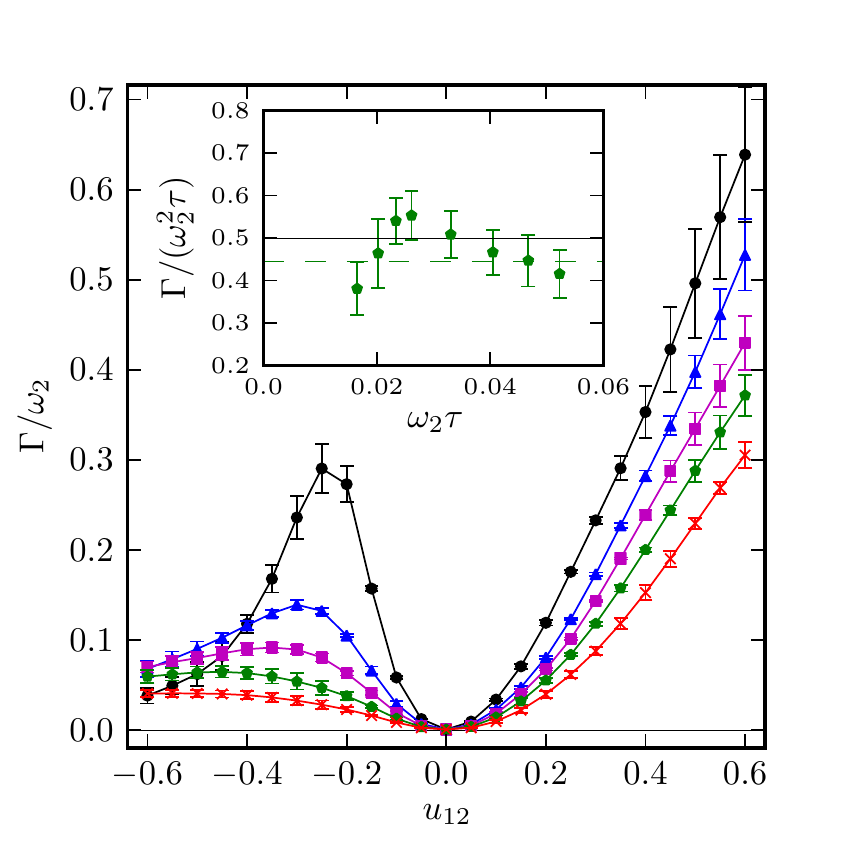} 
\caption{(Color online) Same as in Fig.~\ref{fig:two}, but for the
  friction coefficient $\Gamma$ (in units of $\omega_2$). Note that
  $\Gamma$ vanishes quadratically for weak impurity-bath couplings
  ($|u_{12}| \ll 1$) and saturates to a finite value in the limit $u_1
  \to \infty$.  The inset illustrates the dependence of $\Gamma$ [in
  units of $\omega^2_2 \tau$] on $\omega_2$ (in units of
  $1/\tau$). For each value of $\omega_2$, the tDMRG data (green symbols)
  have been obtained by performing a quench corresponding to
  a value of $\omega_1=\sqrt{10}~\omega_2$.  The
  other parameters are: $N_{\rm bath} = 40$, $L = 600$ ($\langle {\hat
    n}_i\rangle \approx 0.07$), $u_{12} = 0.1$, and $u_1 =1$. The
  solid line represents the prediction $\Gamma_{\rm AL}(\omega_2\ll
  1/\tau) /(\omega^2_2 \tau) = 1/2$, based on the Abraham-Lorentz
  model with the value of $\tau$ corresponding to $u_1 = 1$.  The
  dashed line at $\Gamma/\omega_2^2 \tau = 0.44$ is the result of a
  best fit to the data.\label{fig:three}}
\end{center}
\end{figure}

Fig.~\ref{fig:three} illustrates the dependence of the damping rate
$\Gamma$ on $u_{12}$, for different values of $u_1$. Three features of
the data are remarkable: i) $\Gamma$ displays a strong asymmetrical
behavior with respect to $u_{12}=0$ away from the weak-coupling limit;
ii) $\Gamma$ decreases with increasing $u_1$, saturating to a finite
result in the TG limit; and, finally, iii) $\Gamma$ depends
quadratically on $u_{12}$ for $|u_{12}|\ll 1$.  The non-monotonic
behavior of $\Gamma$ on the attractive side can be explained as
following. For $u_{12} \approx 0$ the damping rate must be small.
Increasing $|u_{12}|$ the damping rate increases because the coupling
of the impurity to the bath increases.  However, upon further
increasing $|u_{12}|$ another effect kicks in. We have indeed
discovered (data not shown here) that $\Gamma$ decreases monotonically
with decreasing frequency. As shown in Fig.~\ref{fig:two}, on the
attractive side $\Omega$ decreases rapidly as $|u_{12}|$ increases,
thereby reducing the damping rate.  The non-monotonic behavior of
$\Gamma$ does not occur for $u_{12} >0$ because on the repulsive side
$\Omega$ changes slightly with respect to $u_{12}$.

\begin{figure}[t]
\begin{center}
\includegraphics[width=1.0\linewidth]{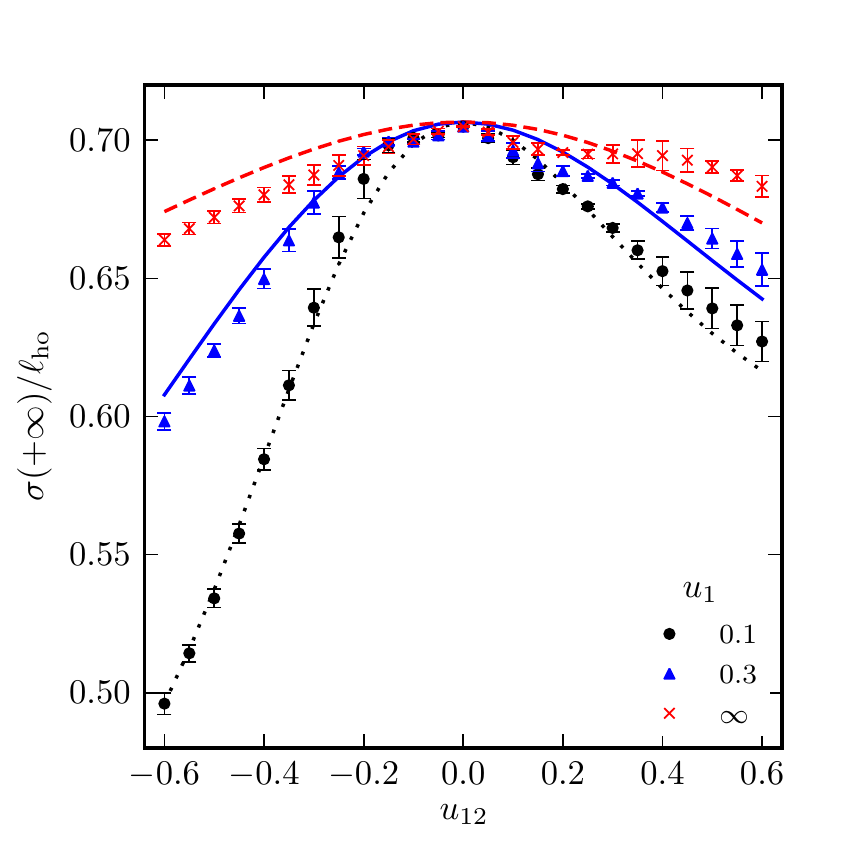} 
\caption{(Color online) Same as in Fig.~\ref{fig:two}, but for the
  asymptotic value $\sigma(+\infty)$ at long times of the width
  $\sigma(t)$ (in units of $\ell_{\rm ho}$). Here the lines are {\it
    not} guides to the eye, but represent the equilibrium value
  $\sigma_{\rm eq}$ for the width in the harmonic potential with
  frequency $\omega_2$ ($\sigma_{\rm eq} = \ell_{\rm ho}/\sqrt{2}$ for
  $u_{12} =0$).\label{fig:four}}
\end{center}
\end{figure}

The asymptotic width $\sigma(+\infty)$, shown in Fig.~\ref{fig:four},
fairly agrees with the equilibrium value at the frequency $\omega_2$
that one can calculate numerically. This finding seems to
suggest that the impurity has nearly ``thermalized" with the bath over
the time scale of our simulations.

\section{Luttinger-liquid theory and the Abraham-Lorentz friction}
\label{sec:ll_al}
We now discuss which features in Figs.~\ref{fig:two}-\ref{fig:four} can
(or cannot) be explained by employing a low-energy Luttinger-liquid
description of the bath.

The Hamiltonian of a single impurity of mass $M$, described by the
pair of conjugate variables (${\hat X}, {\hat P}$), coupled to a bath
of harmonic oscillators (the bosonic excitations of the Luttinger
liquid) with dispersion $\omega_k = v_{\rm s} |k|$,
is~\cite{weiss,Minardi:2012}:
\be\label{eq:ll} 
{\hat {\cal H}}(t) = \frac{{\hat P}^2}{2M} + V({\hat X},
t) + \sum_{k\neq 0} \hbar \omega_k {\hat \gamma}^\dagger_k {\hat
  \gamma}_k + g_{12}{\hat \rho}({\hat X})~, 
\ee
where $V(x,t) = M\omega^2(t)x^2/2$  (the identification with the lattice model fixes $M =
\hbar^2/(2J_2\delta^2)$ with $\delta$ the lattice spacing) and $g_{12}$ is a coupling constant
playing the role of $u_{12}$ in the discrete model.  In
Eq.~(\ref{eq:ll}) ${\hat \gamma}_k^{\dagger}$ (${\hat \gamma}_k$) is
the creation (annihilation) operator for an acoustic-phonon mode with
wave number $k$ and ${\hat \rho}(x) $ is the bath density
operator. The sound velocity $v_{\rm s}$ is related to the Luttinger parameter $K$ 
of the Lieb-Liniger model by Galilean invariance~\cite{haldane:1981}; $K$ is in turn defined by the relation 
$\kappa = K/(\pi \hbar v_{\rm s})$, where $\kappa$ is the compressibility of the bath~\cite{giamarchi}.
The parameters $K$ and $v_s$, which completely characterize the Luttinger liquid, 
can be expressed in terms of the coupling constants of the model in Eq.~(\ref{eq:bath}) (see Sec.~\ref{appendix:luttinger_impurity}). 
We now observe that the sign of the impurity-bath coupling $g_{12}$ can be gauged away
from the Hamiltonian (\ref{eq:ll}) by the canonical transformation
${\hat \gamma}_k \to - {\hat \gamma}_k$. This means that if the bath
was truly a Luttinger liquid, impurity-related observables such as
$\Omega$ and $\Gamma$ should {\it not} depend on $g_{12}$ being
attractive or repulsive. This low-energy description seems to apply
only in a tiny region around $u_{12} =0$. All the deviations from
this prediction seen in Figs.~\ref{fig:two}-\ref{fig:four} have to be
attributed to physics beyond the Luttinger-liquid paradigm.

The Heisenberg equation of motion induced by the Hamiltonian
(\ref{eq:ll}) reads (see Sec.~\ref{appendix:luttinger_impurity}):
\ber\label{eq:llkernel}
M\partial^2_t{\hat X}(t)&+& M \omega^2_2{\hat X}(t)  + M \int_0^t dt'~{\hat \Gamma}(t,t')\partial_{t'}{\hat X}(t') \nonumber\\
&=& - g_{12} \partial_x {\hat \rho(x,t)}\big|_{x = {\hat X}(t)}~, \eer
where 
\be
\label{eq:memorykernel} 
{\hat \Gamma}(t,t') = \sum_{k\neq 0}
\frac{c^2_k k^2}{M \omega^2_k} e^{ik{\hat X}(t)}e^{-ik{\hat X(t')}}
\cos{[\omega_k (t-t')]}~, 
\ee 
with $c_k = -g_{12}[K v_{\rm s}/(\pi \hbar L)]^{1/2}|k|e^{-|k|/2k_c}$,
is the memory kernel~\cite{weiss} ($k_{\rm c}$ is an ultraviolet
cut-off). 

If the dynamics of the impurity is slow with respect to the speed
$v_{\rm s}$ of propagation of information in the bath, then
``retardation effects" can be neglected and we can approximate the
operator ${\hat \Gamma}(t,t')$ with the following c-number
$\Gamma(t-t') = \sum_{k\neq 0} c^2_k k^2\cos{[\omega_k
  (t-t')]}/(M\omega^2_k)$. In this limit it is possible to
show (see Sec.~\ref{appendix:luttinger_impurity}) that Eq.~(\ref{eq:llkernel}) reduces to a quantum
Langevin equation with an {\it Abraham-Lorentz} (AL) term, which
describes the reactive effects of the
emission of radiation from an oscillator~\cite{jackson}. This is a
term of the form $- M \tau \partial_t^3 {\hat X}(t)$ with a
``characteristic time"
%
$\tau  = g_{12}^2K/(\pi M \hbar v_s^4)$.
Remarkably, neglecting the well-known ``runaway"
solution~\cite{jackson} and keeping only the damped solutions, we find
that the quantum Langevin equation with the AL term yields an
expression for $\sigma(t)$ after the quench which is identical to
Eq.~(\ref{eq:motion_final}) with
$
\Gamma = \Gamma_{\rm AL}(\omega_2) \stackrel{\omega_2\tau \ll 1}{=}  \omega^2_2\tau/2
$.
The full functional dependence of $\Gamma_{\rm AL}$ on $\omega_2$ is
reported in Sec.~\ref{appendix:quantum_breathing}.  Note that $\Gamma_{\rm AL}$ is
proportional to $g^2_{12}$.  This is in agreement with the TDMRG
results shown in Fig.~\ref{fig:three} in the weak-coupling $|u_{12}|
\to 0$ limit. Moreover, $\Gamma_{\rm AL}$ is proportional to
$K/v^4_{\rm s} \propto K^5$ (in our case $v_{\rm s}\propto K^{-1}$ from Galilean 
invariance~\cite{haldane:1981}) and proportional to $\omega^2_2$. The former statement
implies a fast saturation of the friction coefficient to a finite
value in the TG limit ($K=1$). This is in agreement with the TDMRG data shown
in Fig.~\ref{fig:three}. The quadratic dependence of the damping rate
on $\omega_2$ is also well displayed by the TDMRG data, at least for
$\omega_2 \tau \ll 1$, as shown in the inset to Fig.~\ref{fig:three}.

It is instructive to compare our findings with the experimental data
of Ref.~\onlinecite{Minardi:2012}.  The latter show that, in a sizable range
of interaction strength $u_{12}$, the frequency of the breathing mode
does not vary appreciably while the damping coefficient increases up
to $\Gamma \sim 0.2~\omega_2$.  As shown in Figs.~\ref{fig:two}
and~\ref{fig:three}, we do observe the same behavior  for $u_1>0.2$.  Moreover,
as in the experiment, we do see that the width of the breathing mode
reduces upon increasing $u_{12}$.  A detailed quantitative
comparison with the experiment is, however, not possible at this
stage: i) one notable difference is that our calculations are carried
out at $T=0$, while temperature effects seem to be important in
Ref.~\onlinecite{Minardi:2012}; ii)  furthermore, the impurity
trapping frequency in our calculations is considerably larger than in the
experiment~\cite{timescales}. Extending the current calculations to
take into account these differences lies beyond the scope of this work.

In summary, we have shown that, in the dynamics of an impurity coupled to a 1D bosonic 
bath, a Luttinger-liquid description of the bath is applicable only 
in a very small region of parameter space, where the impurity 
suffers an AL radiation-reaction friction.  Among the 
most striking features we have found, we emphasize the non-monotonic behavior 
of the damping rate and the large renormalization of the oscillation 
frequency for attractive impurity-bath interactions. 
%

\appendix

\section{Lattice to continuum mapping}
\label{appendix:mapping}

\begin{figure}[t]
  \includegraphics[scale = 1.0]{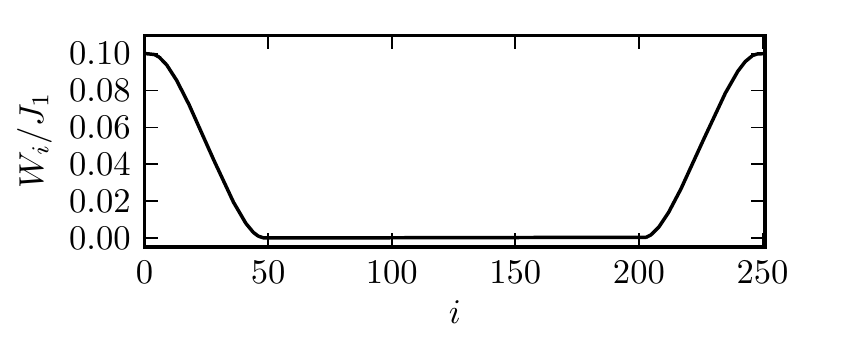}
  \caption{External confining potential for the bath in a lattice with $L=250$ sites, 
    as defined by Eq.~(\ref{eq:external}) (we set $\Delta = 50$ and $W/J_1 = 0.1$).}
  \label{external_potential} 
\end{figure}

In our simulations we consider a 1D bath of interacting bosons coupled to a single impurity 
confined in a harmonic potential. The bath is modeled by a Bose-Hubbard Hamiltonian $\hat{\cal{H}}_{\rm B}$, 
Eq.~(1) in the main text, with hopping $J_1$ and on-site repulsion $U_1>0$.
The external confining potential $W_i$ has the following explicit form:
\begin{equation}
  W_{i} \hspace*{-1mm} = \hspace*{-1mm} \begin{cases}
    W  & \hspace*{-1mm} i = 1\,, \\
    \frac{W}{2} \Big[ 1-\tanh \Big( \frac{3(i-(\Delta+1)/2)}{2\sqrt{(i-1)(\Delta-i)}} \Big) \Big] & \hspace*{-1mm} 1< i< \Delta\,, \\
    0 & \hspace*{-1mm}\Delta \leq i \leq L-\Delta\,, \\
    \frac{W}{2} \Big[ 1+\tanh \Big( \frac{3(i-(2L-\Delta)/2)}{2\sqrt{(i-L+\Delta)(L-i)}} \Big) \Big] &\hspace*{-1mm} L-\Delta< i< L\,, \\
    W & \hspace*{-1mm}i = L\,.
  \end{cases} \label{eq:external}
\end{equation}
The parameter $\Delta$ is the length in unit of lattice sites of the left and right 
boundary regions, where the potential goes from 0 to a finite value $W > 0$. 
The interpolation between the two values is as smooth as possible, since the potential, 
as a continuous function of $x = i\delta$ ($\delta$ being the lattice spacing), 
has zero derivatives of all orders at the joining points in Eq.~(\ref{eq:external}). 
We set $\Delta = 50$ lattice sites, while we used $W/J_1 = 0.1$ for $L = 250$ and $N = 22$ 
(see Fig.~\ref{external_potential}), $W/J_1 = 0.06$ for $L = 600$ and $N = 40$, 
where $L$ is the length of the (two) lattices used in our simulations, 
and $N$ the number of particles in the bath. In the latter case $W$ is smaller 
since the on-site density $n_i \approx 0.07$ in the middle of the chain is smaller and a weaker potential is used. 
%

With our choice of parameters, the local density $\langle\hat n_i \rangle= \langle\hat b_i^\dagger \hat b_i\rangle$ 
is essentially constant in a region of length $\sim 2L/3$, and kept to a low value 
$\langle \hat n_i \rangle \lesssim 0.1$ everywhere. 
At such low densities and for not too strong repulsive interaction ($U_1/J_1 \lesssim 10$), 
the lattice is irrelevant and the model can be mapped to a Lieb-Liniger 
Hamiltonian~\cite{cazalilla_rmp_2011} describing 1D bosons of mass $m = \hbar^2/(2J_1\delta^2)$ 
interacting through a contact (repulsive) two-body potential. 
The upper bound for the parameter $U_1/J_1$ can be understood as follows: 
as the interaction between bosons is increased, the healing length of the Lieb-Liniger gas gets smaller, 
and, when it is comparable with $\delta$, lattice effects becomes relevant. 
An extensive discussion of this point can be found in Refs.~\onlinecite{Cazalilla_pra1, Cazalilla_pra2, Kollath_pra}. 
Other relevant parameters of the continuum model are the density $n = \langle \hat
n_i\rangle/\delta$ (where $\langle \hat n_i\rangle$ is the on-site density
taken in the central region where the bath is homogeneous) and the
dimensionless Lieb-Liniger parameter $\gamma_1 = mg_1/(\hbar^2 n) =
U_1/(2J_1\langle \hat n_i\rangle)$, where $g_1 = U_1\delta >0$ is the
strength of the contact repulsion between bosons.  

The Hamiltonian $\hat{\cal{H}}_{\rm I}(t)$ describing the impurity is written in Eq.~(2) in the main text. 
Given the low impurity density $\langle \hat N_i \rangle \lesssim 0.15$, also in this case 
the lattice model well describes the continuum Hamiltonian of a particle of mass 
$M = \hbar^2/(2J_2\delta^2)$ moving in a parabolic potential 
$V(x,t) = V_2(t) (x/\delta)^2 = M\omega^2(t)x^2/2$ (centered without loss of generality at $x=0$). 
The impurity mass is fixed to $M = m/2$ (thus corresponding to a ratio between the two
lattice hoppings $J_2 / J_1 = 2$), such to take into account the mass imbalance between 
Rb and K atoms, as experimentally done in Ref.~\onlinecite{Minardi:2012}.

The total Hamiltonian 
${\hat {\cal H}}(t) = {\hat {\cal H}}_{\rm B} + {\hat {\cal H}}_{\rm I}(t) + {\hat {\cal H}}_{\rm coupl}$ 
contains a further density-density coupling between bath and impurity, 
${\hat {\cal H}}_{\rm coupl} = U_{12}\sum_i {\hat n}_i{\hat N}_i$, 
which in the continuum limit corresponds to a $\delta$-function of strength $g_{12} = U_{12}\delta$.

\section{Quantum Langevin equation for a particle in a Luttinger liquid}
\label{appendix:luttinger_impurity}

In this section we derive a quantum Langevin equation~\cite{ford_pra_1988, weiss,gardiner} 
for a single particle, described by the conjugate variables $(\hat{X}, \hat{P})$, 
that is coupled to a Luttinger liquid by a density-density interaction. 
The relevant Hamiltonian~\cite{Minardi:2012} is given by Eq.~(\ref{eq:ll})
where the first two terms describe the impurity Hamiltonian, 
the potential $V(\hat{X}) \equiv -M \omega^2 \hat{X}^2/2$ representing an harmonic 
confining trap of frequency $\omega$ for the impurity, 
while the third one denotes the quadratic Luttinger Hamiltonian, 
in which $\hat \gamma_k^{\dagger}$ ($\hat \gamma_k$) is the creation (annihilation) operator for 
an acoustic-phonon mode with wave vector $k$ and dispersion $\omega_k = v_{\rm s}|k|$. 
Apart from the coupling constant $g_{12}$, the last term, which contains 
an ultraviolet cut-off ($k_{\rm c}$), defines the bath density operator $\hat \rho(\hat{X})$ 
of the Luttinger liquid~\cite{giamarchi}, and is controlled by the parameter $K$.
This, in turn, controls the speed of ``sound'' $v_{\rm s}$ by virtue of Galilean invariance: 
$v_{\rm s} = \hbar\pi n/(m K)$. Repulsive interactions in the bath enter the problem 
through the dependence of $v_{\rm s}$ and $K$ on the Lieb-Liniger parameter $\gamma_1$~\cite{cazalilla_rmp_2011}.

In Eq.~(\ref{eq:ll}) the phonon modes couple linearly to the particle position, 
the latter entering through the exponential $\exp{(ik\hat X)}$. 
As a consequence, the harmonic excitations of the Luttinger liquid can be integrated out leaving an effective 
dissipative equation (quantum Langevin equation~\cite{ford_pra_1988}) for the impurity degree of freedom.
We first switch from annihilation and creation operators to (complex) position and momentum, by defining 
${\hat x}_k = \sqrt{\hbar/2\omega_k}\big({\hat \gamma}_k+{\hat \gamma}_{-k}^{\dagger}\big)$ and 
${\hat p}_k =i\sqrt{\hbar \omega_k/2}\big({\hat \gamma}_k^\dagger-{\hat \gamma}_{-k}\big)$, 
with $\big[\hat x_k, \hat p_{k'}\big]  = i\hbar\delta_{kk'}$. 
Notice that, unlike the usual position and momentum, they are complex 
and their adjoint are ${\hat x}^\dagger_{k} = {\hat x}_{-k}$ and ${\hat p}^\dagger_{k} = -{\hat p}_{-k}$.
Using these new variables one gets 
\begin{eqnarray}\nonumber
    \hat{\mathcal H} & = & \frac{\hat P^2}{2M} + V(\hat X) 
    + \sum_{k>0} \bigg( |{\hat p}_k|^2+\omega_k^2\Big| {\hat x}_k-\frac{c_k}{\omega_k^2}e^{-ik{\hat X}}\Big|^2 \bigg)\\ 
    & + & \text{const.} \,, \label{eq:ll2}
\end{eqnarray}
where the coefficients $c_k = -g_{12} \sqrt{\frac{Kv_{\rm s}}{\pi \hbar L}}|k|e^{-|k|/2k_c}$ 
have been defined according to Ref.~\onlinecite{weiss}. Usually, after completing the square, 
some additional terms depending only on $\hat X$ are left out and their effect is to renormalize 
the potential $V(\hat X)$. 
Given the peculiar non-linear nature of the coupling $\propto \exp({ik\hat X})$, 
this does not happen in our case: the terms completing the square turn out to be independent
of $\hat{X}$, therefore the potential $V(\hat{X})$ is not renormalized.

The Heisenberg equations of motion dictated by Eq.~(\ref{eq:ll2}) are given by
\begin{eqnarray}
  M\partial_t^2\hat X -F({\hat X}) & = & \sum_{k \neq 0} i k c_k {\hat x}_{k} e^{ik{\hat X}} \,, \label{eq:heisenberg1}\\
  \partial_t^2{\hat x}_k + \omega_k^2 {\hat x}_k & = & c_ke^{-ik{\hat X}} \, ,
\end{eqnarray}
where $F(x) \equiv -\partial_xV(x)$.
The solution of the second equation can be immediately written down:
\begin{eqnarray}
  {\hat x}_k(t) & = & {\hat x}_k(0) \cos \omega_k t + \frac{{\hat p}_{-k}(0)}{\omega_k} \sin \omega_k t \nonumber \\
  & + & \frac{c_k}{\omega_k}\int_0^tdt' \, \sin [\omega_k(t-t')] \, e^{-ik{\hat X}(t')} \, .
\end{eqnarray}
Integrating by parts and substituting into Eq.~(\ref{eq:heisenberg1}), we get
\begin{widetext}
  \begin{equation}\label{eq:rad_damping_first}
    M \partial_t^2{\hat X}(t) - F({\hat X}(t)) + M \int_0^t dt' \, \hat{\Gamma}(t,t') 
    \partial_{t'} {\hat X}(t') = \hspace*{-1mm} 
    \sum_{k\neq0} ikc_ke^{ik{\hat X(t)}} \Big[ \Big( {\hat x}_k(0)-\frac{c_k}{\omega_k^2} e^{-ik{\hat X}(0)} \Big) \cos\omega_k t
      + \frac{{\hat p}_{-k}(0)}{\omega_k}\sin \omega_kt \Big] 
  \end{equation}
\end{widetext}
with a memory kernel function
\begin{equation}
  \hat{\Gamma}(t,t') = \hspace*{-1mm} \sum_{k\neq 0}\frac{c^2_kk^2}{M \omega_k^2}
  e^{ik{\hat X}(t)}e^{-ik{\hat X(t')}} \cos \omega_k(t-t')\,. 
  \label{eq:kernel}
\end{equation}
that depends {\it separately} on $t$ and $t'$ (and not only on $t-t'$), due to the presence 
of the two non-commuting operators ${\hat X}(t)$ and ${\hat X}(t')$ evaluated at different times.
Upon the substitution $\hat x_k(0) -c_ke^{-ik\hat X(0)}/\omega^2_k\to \hat x_{k}(0)$, 
the right hand side is just $-g_{12}\partial_x\hat \rho(x,t)|_{x =\hat X(t)}$. 
In order to see this, simply rewrite $\hat x_k(0)$ and $\hat p_k(0)$ using the corresponding 
annihilation and creation operators and compare to the density-density coupling in the Hamiltonian~(\ref{eq:ll}).
As discussed in Ref.~\onlinecite{weiss}, one can forget about the term 
$\propto e^{ik\hat X(0)}$ in the right hand side of Eq.~(\ref{eq:rad_damping_first}), 
since it has no effect when taking averages on the equilibrium state of the bath 
coupled to the particle at $t = 0$ (as a consequence 
the stochastic term in the quantum Langevin equation~(\ref{eq:rad_damping_first}) 
has zero average: $\langle \partial_x \hat \rho(\hat X(t),t)\rangle = 0$). 

When evaluating the noise correlator one runs into difficulties since 
$e^{ik\hat X(t)}e^{-ik\hat X(t')}\neq e^{ik[\hat X(t)-\hat X(t')]}$.
In the limit when the particle has a small velocity with respect to $v_{\rm s}$, 
we can however perform the following approximation:
\begin{equation} 
  e^{ik\hat X(t)}e^{-ik\hat X(t')}e^{\pm iv_{\rm s}|k|(t-t')} \approx e^{\pm iv_{\rm s}|k|(t-t')} \, ,
  \label{eq:slowpart}
\end{equation}
since in this case the product of the first two exponentials is assumed to be
slowly varying with respect to the third one. 
Therefore in this limit the noise correlator $\hat{\Xi}(t,t') \equiv 
g_{12}^2 \langle[ \partial_x \hat\rho({\hat X}(t),t)]^{\dagger} \partial_x \hat\rho({\hat X}(t'),t')\rangle$ reads:
\begin{equation} \label{eq:noise_corr}
\begin{split}  
\hat{\Xi}(t,t') \approx 
  \frac{g_{12}^2 K}{\pi \hbar v_s^4} \int_{0}^{+\infty} \frac{d\omega}{2\pi} e^{-\omega/\omega_c}\hbar \omega^3 \, \times \\
   \times \left(\coth \frac{\beta\hbar\omega}{2}\cos\omega(t-t')-i\sin\omega(t-t')\right) \,,
\end{split}
\end{equation}
where we passed in the continuum by substituting the series over $k$ with an integral
over the frequencies $\omega$.
In the same approximation, the memory kernel~(\ref{eq:kernel}) entering
the third term in the left hand side of Eq.~(\ref{eq:rad_damping_first}) reads
\begin{equation}
  \hat{\Gamma}(t,t') \approx \frac{2g_{12}^2K}{\pi \hbar v_{\rm s}^4} \int_0^{+\infty} \frac{d\omega}{2\pi} 
  e^{-\omega/\omega_c} \omega^2 \cos\omega(t-t') \,. 
\end{equation}
Notice that, when the cut-off $\omega_c$ goes to infinity, this kernel basically reduces 
to the second derivative of a delta function:
\begin{equation}
  \hat{\Gamma}(t,t') \stackrel{\omega_c\to +\infty}{=}-M\tau \delta''(t-t') \,,
\end{equation}
where $\tau = g_{12}^2K/(\pi M\hbar v_{\rm s}^4)$ is the proper time scale.

The relation 
$\Re \big[\widetilde\Xi(\omega)\big] = M\hbar \omega\coth(\beta\hbar\omega/2)\Re \big[\widetilde\Gamma(\omega)\big]$
states the fluctuation-dissipation theorem~\cite{ford_pra_1988,weiss,gardiner}, and 
holds between the real parts of the Fourier transforms of the noise correlator, $\widetilde{\Xi}(\omega)$,
and of the memory kernel, $\widetilde{\Gamma}(\omega)$. 
This ensures the consistency of the {\it slow-particle approximation}, see Eq.~(\ref{eq:slowpart}). 
Within this approximation, the quantum Langevin equation takes the linear form 
\begin{equation} \label{eq:lin_rad_damping}
M\partial_t^2{\hat X}(t) -F({\hat X}(t)) -M\tau \partial_t^3{\hat X}(t) = -g_{12}\partial_x\hat \rho(\hat X(t), t)\,.
\end{equation}
The fact that the position operator appears on the right hand side as an argument 
of the bath density does not spoil linearity, since the noise correlator is independent 
of the difference $X(t)-X(t')$ in the slow-particle approximation.

Considering $\hat X(t)$ as a {\it classical} variable, the $k$-sum in the third term on 
the left of Eq.~(\ref{eq:rad_damping_first}) can be easily performed, thus obtaining
the following classical Langevin equation:
\begin{equation}
\begin{split} \label{eq:rad_damping}
  &M \partial_t^2X(t) - F({X}(t)) -\frac{g_{12}^2K}{2\pi \hbar v^3_{\rm s}} 
  \int_0^tdt' \, \partial_t^3{X}(t') \, \times \\ &\times \sum_{\epsilon = \pm}
  \big[ \delta \big( {X}(t) - {X(t')} + \epsilon \, v_{\rm s}(t-t') \big) \big] = -g_{12}\partial_{x}{\rho}({X}(t), t)\,.
\end{split}
\end{equation}
From here it is apparent that the particle is interacting at point $X(t)$ 
and time $t$ with the phonons (density fluctuations in the Luttinger liquid) 
emitted by itself at point $X(t')$ in a past time $t'$. 
This is similar to radiation damping of the motion of a charge particle, where the role 
of the electromagnetic field is now played by the Luttinger liquid. 
The problem of damping due to the emission of radiation is a very old one 
and the classical version of Eq.~(\ref{eq:lin_rad_damping}) has been known in this context 
for a long time as the Abraham-Lorentz equation~\cite{jackson}. 

The quantum version of Eq.~(\ref{eq:rad_damping}) is more complicated, 
due to its operatorial character (in this case indeed $e^{ik\hat X(t)}e^{-ik\hat X(t')}\neq e^{ik[\hat X(t)-\hat X(t')]}$).
In the specific, Eq.~(\ref{eq:lin_rad_damping}) has been obtained under the assumption~(\ref{eq:slowpart}),
thus it is expected to be valid only when the particle does not move too fast with respect to the phonons.
This can be quantified by considering a small value for the ratio $\nu$ between the maximum velocity 
of the impurity distribution when it expands, $\omega_2\ell_{\rm ho}^2/\sigma(0)$
(here $\ell_{\rm ho} = \sqrt{\hbar/M\omega_2}$ is the harmonic oscillator length),
and the velocity of sound $v_s$.
Such ratio is independent of the trap frequency, and is fixed only by the initial squeezing 
through the uncertainty principle -- if the uncertainty in position is $\sigma(0)$ 
then the uncertainty in velocity is $\hbar /(M \sigma(0))$. 

In our simulations this parameter is actually not so small. 
An estimate can be given as follows: the velocity of sound $v_{\rm s}$ is a fraction 
of the Fermi velocity for a gas of free fermions at the same density, 
say $v_{\rm s} \sim \hbar \pi n/m$, the initial squeezing is $\sigma(0)/\ell_{\rm ho} \approx 0.4$ 
and the density $n\ell_{\rm ho} \approx 1$, so 
\begin{equation}
  \nu = \frac{\omega_2\ell_{\rm ho}^2}{v_s \sigma(0)} 
      = \frac{\hbar} {M v_{\rm s} \sigma(0)} = \frac{m/M}{\pi \sigma(0)n}  \approx 1.6 \, .
\end{equation}
So we expect corrections to Eq.~(\ref{eq:lin_rad_damping}) due to retardation effects 
embodied in Eq.~(\ref{eq:rad_damping_first}).

\section{Impurity breathing mode within the quantum Langevin equation}
\label{appendix:quantum_breathing}

In this section we derive the function used to fit the TDMRG data [see Eq.~(3) in the main text].
We show how it can be obtained starting both from the quantum Langevin equation 
for an {\it ohmic} bath~\cite{weiss,gardiner}:
\begin{equation}\label{eq:langevin_ohmic}
  M \partial^2_t {\hat X}(t)+ 2 M \Gamma \partial_t {\hat X}(t) + M\omega^2_2{\hat X}(t) = {\hat \xi}(t)\,,
\end{equation}
with noise correlator
\begin{eqnarray}
  \Xi(t) & = & \langle {\hat\xi}(t){\hat\xi}(0)\rangle\\
         & = & \frac{2M\Gamma}{\pi} \int_{0}^{+\infty} d\omega \hbar \omega
               \left[\coth\frac{\beta\hbar\omega}{2}\cos\omega t -i\sin\omega t\right] \, , \nonumber
\end{eqnarray}
and from the Langevin equation~(\ref{eq:lin_rad_damping}) in the previous section,
with a noise correlator given by Eq.~(\ref{eq:noise_corr})
that describes a {\it superohmic} bath~\cite{weiss,gardiner}. 

The predictions for the oscillation frequency and the damping coefficient are different in the two cases. 
However it is possible to treat both equations in the same way, because their respective 
Green functions $G(\omega)$ have a similar pole structure in the frequency domain. 
Both have two poles at complex frequencies $\omega_{\pm}$ with negative imaginary part ($\Gamma > 0$):
\begin{equation}\label{eq:roots}
  \omega_{\pm} = \pm \sqrt{\Omega^2-\Gamma^2} -i\Gamma\,.
\end{equation} 
These are the physically relevant ones and  correspond to exponentially decaying solutions. 
It is essential for our purposes that $\omega_+ = -\omega_-^*$. 
We point out that Eq.~(\ref{eq:lin_rad_damping}) is of third order and its Green function has a third pole 
in the upper half of the complex plane that corresponds to an unphysical diverging solution, 
also called ``run-away'' solution~\cite{jackson}, which will be discarded in the following.
Note that, while for the ohmic case $\Omega$ is equal to the trap frequency $\omega_2$, 
and $\Gamma$ is exactly the coefficient 
that appears in Eq.~(\ref{eq:langevin_ohmic}), for the Langevin equation in~(\ref{eq:lin_rad_damping}) 
one has a more complex dependence on the trap frequency $\omega_2$ and on the time scale $\tau$. 
In the specific, the three roots of the {\it cubic} characteristic 
equation $-i\omega^3\tau-\omega^2+\omega_2^2=0$ read
\begin{gather} \label{eq:omegapm}
    \omega_{\pm} = \pm \frac{1}{2\sqrt{3}\tau}\left(z-z^{-1}\right)+\frac{i}{3\tau}\left(1-\frac{z+z^{-1}}{2}\right) \,, \\
  \omega_{\rm run-away} = \frac{i}{3\tau}\left[1+z+z^{-1}\right]\,,
\end{gather}
with
\begin{equation}
  z^{\pm 1} = \bigg( \frac{27\omega_2^2 \tau^2 + 2\pm\sqrt{(27\omega_2^2\tau^2+2)^2-4}}{2} \bigg)^{1/3} \, .
\end{equation}
Note that $\omega_{\pm}$ have always negative imaginary part, 
while $\omega_{\rm run-away}$ has positive imaginary part and zero real part. 

The symmetry of the physical roots allows us to write the solution of both the two Langevin equations as
\begin{equation}\label{eq:sol}
  \begin{split}
    \hat X(t) &= \hat X(0)e^{-\Gamma t}\cos\sqrt{\Omega^2-\Gamma^2} t \\
    &\quad+\frac{\partial_t\hat{X}(0)+\Gamma\hat X(0)}{\sqrt{\Omega^2-\Gamma^2}}e^{-\Gamma t}\sin \sqrt{\Omega^2-\Gamma^2} t \\
    &\quad+\int_0^{t}dt'G(t-t')\hat \xi(t') \, ,
  \end{split}
\end{equation}
$\hat \xi(t)$ and $G(t)$ denoting the noise term and the Green function respectively. 

Using Eq.~\eqref{eq:sol}, and the fact that $\langle\hat \xi(t)\rangle = 0$,
we can evaluate some asymptotic averages which turn out to be useful to calculate
the impurity breathing mode. Some of them are null:
\begin{equation}
  \langle \hat X(+\infty)\rangle = \langle \partial_t \hat X(+\infty)\rangle  = 
  \langle \{\hat X,\partial_t \hat X \}\rangle_{t \to +\infty} = 0 \, ,
  \label{eq:avgzero1}
\end{equation}
while other ones, such as $\langle \hat X^2(+\infty) \rangle$ and $\langle (\partial_t \hat X(+\infty))^2\rangle$,
can be implicitly written in terms of the spectral function~\cite{weiss}, whose form depends on the nature of the bath.
Finally, call $\hat E(t) = \int_0^{t}dt'G(t-t')\hat \xi(t')$ and note that
\begin{equation}\label{eq:xve_mean}
  \langle \hat X(0)\hat E(t)\rangle = \langle \partial_t \hat X(0) \hat E(t)\rangle = 0\,.
\end{equation}

With all these results at hand, the average of $\hat X^2(t)$ can be calculated at an arbitrary time $t$.
Suppose that the particle starts its motion at $t = -\infty$ in a squeezed harmonic potential. 
At $t=0$ it will have equilibrated with the bath, and expectation values are taken on the 
equilibrium state of the {\it whole} system, with the particle at rest in the squeezed harmonic potential. 
Then the frequency of the harmonic confinement is suddenly quenched to a new value $\omega_2$. 
The width of the impurity density distribution as a function of time, for $t>0$, 
can be explicitly calculated by taking the square of Eq.~(\ref{eq:sol}) on the global state:
$\sigma^2(t) = \langle \hat X^2(t)\rangle$.
Using Eqs.~(\ref{eq:avgzero1})-\eqref{eq:xve_mean}, one ends up with the following expression:
\begin{multline}\label{eq:motion}
    \sigma^2(t) = \langle \hat E^2(t) \rangle + \frac{e^{-2\Gamma t}}{\cos^2\phi} 
    \bigg[ \langle \hat X^2(0) \rangle \cos^2 \big( \sqrt{\Omega^2-\Gamma^2}t-\phi \big) \\ 
      + \frac{\langle (\partial_t \hat X(0))^2\rangle}{\Omega^2}\sin^2 \sqrt{\Omega^2-\Gamma^2}t \bigg] \,,
\end{multline}
with $\phi = \arcsin (\Gamma/\Omega)$.
It is not necessary to calculate explicitly the average $\langle \hat E^2(t)\rangle$, 
since it has to cancel exactly the other terms in Eq.~\eqref{eq:motion} 
when $\langle \hat X^2(0)\rangle = \langle \hat X^2(+\infty)\rangle$ and 
$\langle (\partial_t \hat X(0))^2\rangle = \langle (\partial_t \hat X(+\infty))^2\rangle$. 
Therefore the final result is
\begin{multline}\label{eq:motion2}
  \sigma^2(t) = \sigma^2(+\infty) \bigg\{ 1 + \frac{e^{-2\Gamma t}}{\cos^2\phi}
                \Big[ \Delta_x \cos^2 \big( \sqrt{\Omega^2-\Gamma^2}t-\phi \big) \\
                   + \Delta_p \sin^2 \sqrt{\Omega^2-\Gamma^2}t \Big] \bigg\} \, ,
\end{multline}
where the definitions below have been used:
\begin{gather}\label{eq:x1}
  \Delta_x = \frac{\langle \hat X^2(0)\rangle-\langle \hat X^2(+\infty)\rangle}{\langle \hat X^2(+\infty)\rangle}, \\
  \Delta_p = \frac{\langle (\partial_t \hat X(0))^2\rangle-\langle (\partial_t \hat X(+\infty))^2\rangle}
                   {\Omega^2\langle \hat X^2(+\infty)\rangle}\,.
\end{gather}

We employ Eq.~\eqref{eq:motion2} in our fitting procedure as follows. 
$\Delta_x$ and $\Delta_p$ are fixed to their non-interacting values
\begin{equation}
  \Delta_x = \frac{\sigma^2(0)}{\sigma^2(+\infty)}-1\,\quad\text{and}\quad
  \Delta_p = \frac{\sigma^2(+\infty)}{\sigma^2(0)}-1\,,
\end{equation} 
while $\sigma(+\infty)$, $\Omega$, $\Gamma$ are used as fitting parameters. 
The initial width $\sigma(0)$ is provided directly by the numerical data. 
This choice allows reliable fits that smoothly interpolate with the exact non-interacting result:
\begin{equation}
  \sigma(t) = \sqrt{\sigma^2(0)\cos^2\omega_2t + \frac{\ell_{\rm ho}^4}{4\sigma^2(0)}\sin^2\omega_2t}\,.
\end{equation}

\begin{figure*}[!t]
  \includegraphics[scale = 1.0]{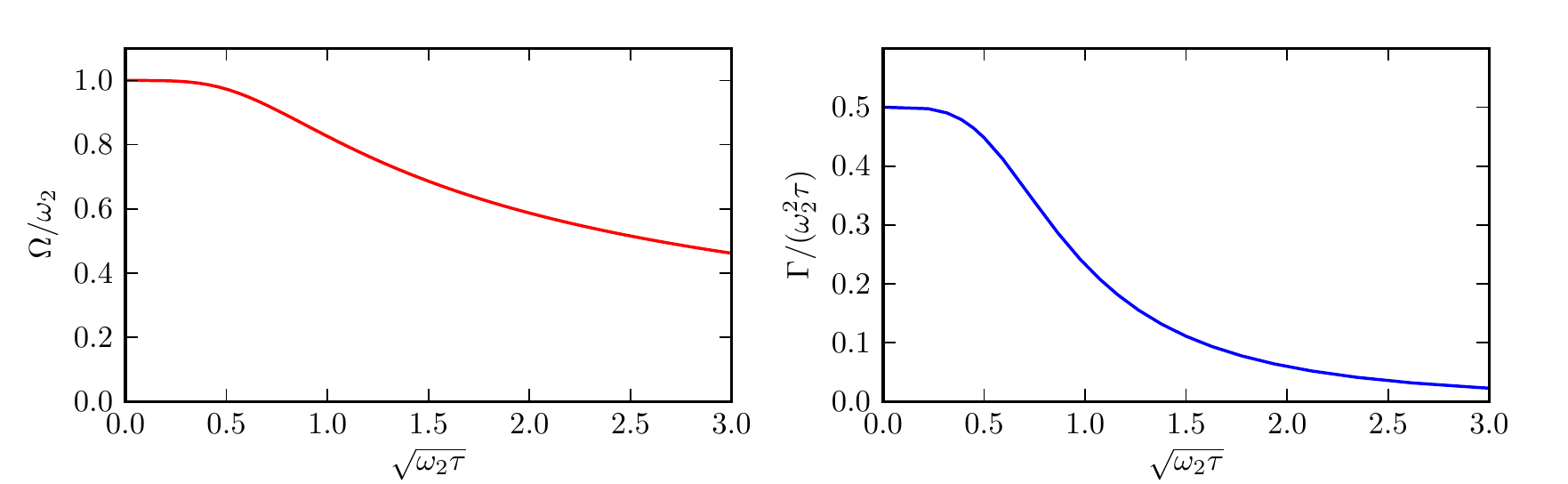}
  \caption{(Color online) Plot of $\Omega/\omega_2$ and $\Gamma/\omega_2$ 
    as a function of $\sqrt{\omega_2\tau}\propto g_{12}$, to allow a comparison with the results in the main text.
    In particular, after expanding Eq.~\eqref{eq:omegapm} for $\omega_2 \tau \ll 1$,
    we get: $\Omega /\omega_2 \approx 1 - \omega_2^2 \tau^2 /2$ and $\Gamma / (\omega_2^2 \tau) \approx 1/2$.}
  \label{varpi-gamma}
\end{figure*}

Surprisingly, the same functional form of Eq.~(\ref{eq:motion2}) holds for two quite different models.
This is due to the form of the two relevant complex frequencies [see Eq.~(\ref{eq:roots})] 
that applies to both of them. However the two models provide distinct predictions 
for the modulus $\Omega = |\omega_{\pm}|$ and the damping coefficient $\Gamma$,
with some appreciable differences.
For an ohmic bath one has $\Omega = \omega_2$, so that the frequency 
is not renormalized in this case. Moreover $\Gamma$ is a fixed time constant 
independent of the trap frequency. 
For the superohmic case of Eq.~(\ref{eq:lin_rad_damping}), for $\omega_2\tau\ll 1$ it can be shown 
that $\Omega/\omega_2$ is a function of $\omega_2\tau$, the same holds for $\Gamma/\omega_2$ (see Fig.~\ref{varpi-gamma}). 
Notice also that $\omega_{\pm}$ always has a non-zero real part,
therefore there is no overdamped solution for any value of $\tau$.

\begin{figure*}[!h]
  \includegraphics[width=1.0\linewidth]{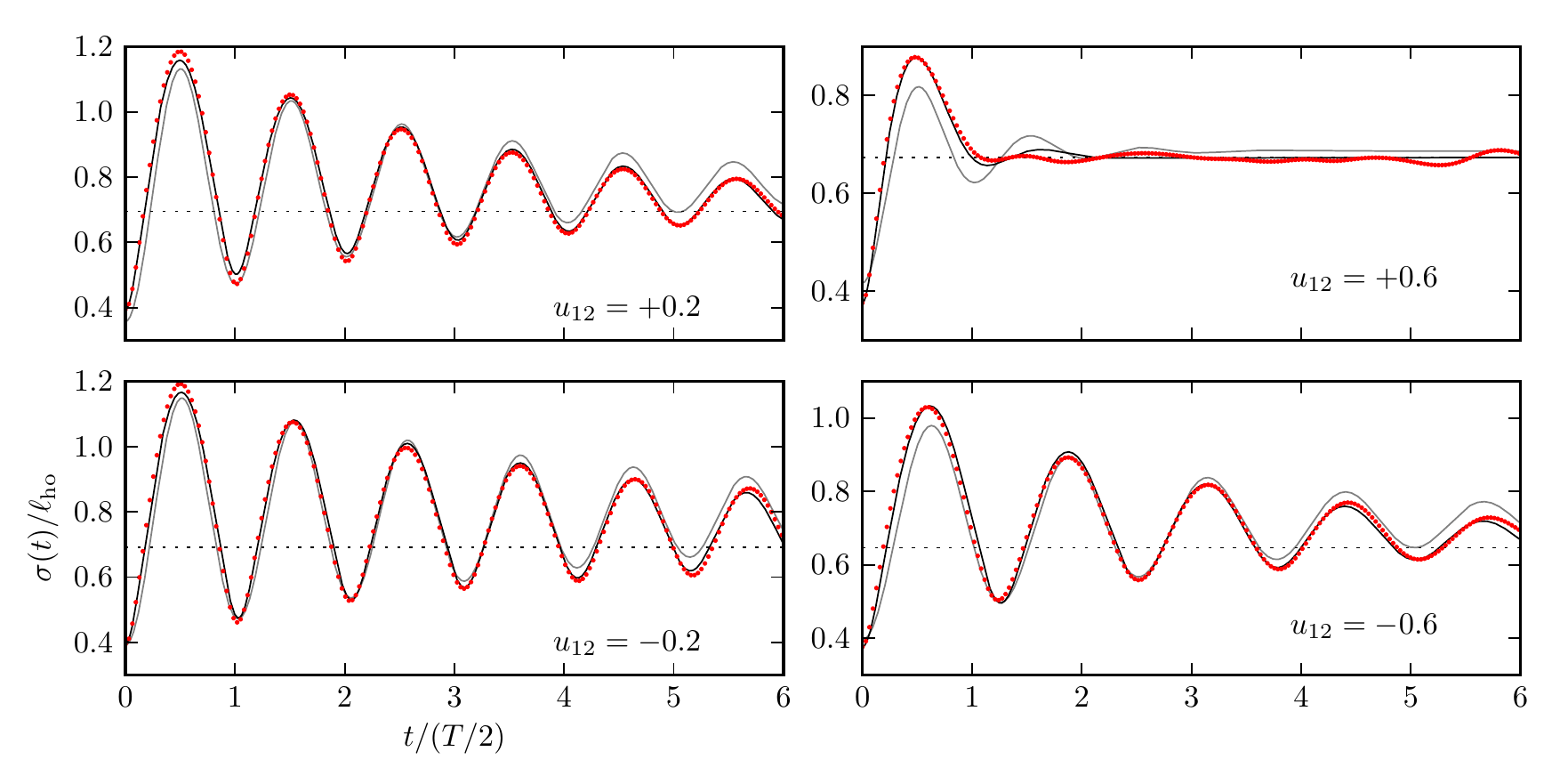} 
  \caption{(Color online) Width of the impurity breathing mode $\sigma(t)$ (in units of $\ell_{\rm ho}
    = [\hbar/(M \omega_2)]^{1/2}$) as a function of time $t$ (in units of $T/2 = \pi/\omega_2$). 
    Data in different panels corresponds to four values of the impurity-bath Lieb-Liniger 
    parameter $u_{12}$, for a fixed value of the bath Lieb-Liniger parameter $u_1 = 1$. 
    Filled red circles label the tDMRG results. The black solid lines are fits to the tDMRG data
    using Eq.~(\ref{eq:motion2}). The thin grey lines are obtained by employing 
    the fitting function in Eq.~(\ref{eq:naive}).}
  \label{fig:two_bis}
\end{figure*}

Finally we emphasize that the functional form given in Eq.~\eqref{eq:motion2} is essential 
to extract the parameters $\Omega$, $\Gamma$ and $\sigma(+\infty)$ from the numerical data. 
Indeed we have verified that the numerical fits performed using the functional form~\cite{Minardi:2012}
\begin{equation}\label{eq:naive}
  \sigma(t) = \sigma(+\infty)+Ae^{-\Gamma t}\cos( \sqrt{\Omega^2-\Gamma^2} t-\phi)\,,
\end{equation}
with $A$ being the oscillation amplitude (grey lines in Fig.~\ref{fig:two_bis}), are significantly different 
from the ones that are obtained employing Eq.~(\ref{eq:motion2}) (black lines in Fig.~\ref{fig:two_bis}). 
For example, the fit done with Eq.~(\ref{eq:naive}) often overestimates the asymptotic with $\sigma(+\infty)$.

\acknowledgements We acknowledge financial support by the EU FP7 Grant
No. 248629-SOLID.


\begin{thebibliography}{77}
\bibitem{footnotetime} 
  Here $\langle {\hat N}_i(t)\rangle$ is a shorthand for 
  $\langle\Psi(t)|{\hat N}_i|\Psi(t)\rangle$, $|\Psi(t)\rangle$ being 
  the state of the system at time $t$.
\bibitem{initialcondition}
  The initial condition is given by the ground state of $\hat {\cal H}(t)$ at times $t<0$.
\end{thebibliography}
\end{document}